\documentclass[]{jfm}
\usepackage{graphicx,natbib,upmath}
\usepackage{amssymb}

\title[Channelization in Permeable Ground]{
Spontaneous Channelization in Permeable Ground:
Theory, Experiment, and Observation}

\author[N. Schorghofer, B. Jensen, A. Kudrolli, and D.H. Rothman]{
N\ls O\ls R\ls B\ls E\ls R\ls T\ns 
S\ls C\ls H\ls O\ls R\ls G\ls H\ls O\ls F\ls E\ls R$^1$,\ns
B\ls I\ls L\ls L\ns J\ls E\ls N\ls S\ls E\ls N$^{2,3}$,\ns\\
A\ls R\ls S\ls H\ls A\ls D\ns 
K\ls U\ls D\ls R\ls O\ls L\ls L\ls I$^2$\ns 
\and D\ls A\ls N\ls I\ls E\ls L\ns H.\ns 
R\ls O\ls T\ls H\ls M\ls A\ls N$^1$}

\affiliation{$^1$Department of Earth, Atmospheric and Planetary Sciences,\\
Massachusetts Institute of Technology, Cambridge, MA 02139, USA\\[\affilskip]
$^2$Department of Physics, Clark University, Worcester, MA 01610, USA\\[\affilskip]
$^3$Department of Physics, University of Massachusetts, Boston, MA 02125, USA}

\date{24 September 2003}

\begin{document}
\maketitle

\begin{abstract}
Landscapes that are rhythmically dissected by natural drainage channels
exist in various geologic
and climatic settings.  Such landscapes are characterized by a
length-scale for the lateral spacing between channels.  We observe a
small-scale version of this process in the form of beach rills and
reproduce channelization in a table-top seepage experiment.  On the
beach as well as in the experiments, channels are spontaneously
incised by surface flow, but once initiated, they grow due to water
emerging from underground.  Field observation and experiment suggest
the process can be described in terms of flow through a homogeneous
porous medium with a freely shaped water table.  According to this
theory, small deformations of the underground water table amplify the
flux into the channel and lead to further growth, a phenomenon we call
``Wentworth instability''.  Piracy of groundwater can occur over
distances much larger than the channel width.  Channel spacing
coarsens with time, until channels reach their maximum length.
\end{abstract}


\section{Introduction}

\begin{figure}
\centerline{\includegraphics[width=9cm]{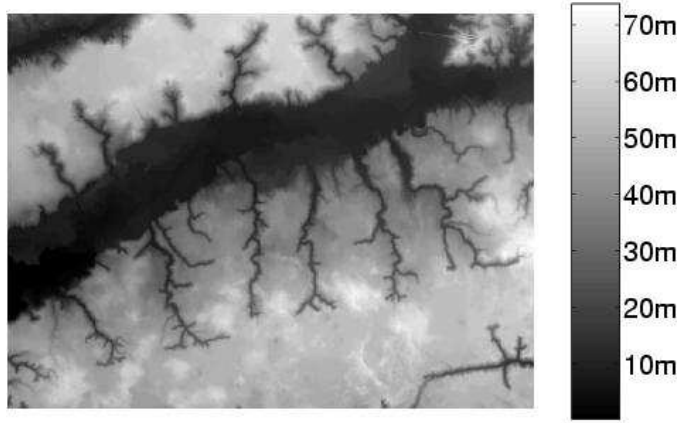}}
\caption{Elevation map of sapping canyons, Okaloosa, Florida
\citep{sbws95}.  This is one example of periodically spaced channels
in nature.  The image is 36km wide and the spacing is roughly 5km.
\label{firstexample}}
\end{figure}

Nature frequently exhibits regularly spaced channels that are formed by
fluvial erosion.  Figure~\ref{firstexample} shows one such example
with approximately regularly spaced drainage outlets.  The spacing in
naturally occurring examples varies from tens of centimeters to tens
of kilometers, spanning five orders of magnitude.  Some examples are
more nearly periodic than others, but in either case there is a
typical length-scale for the lateral separation of streams.  This
length-scale is of interest not only because it determines the shape
of landforms around us, but also because it sets one of the scales of
variation in the heterogeneities of sedimentary deposits
\citep{talling97}.

The ubiquity of this phenomenon further motivates its study.
Periodically spaced drainage outlets are observed in various climates
and geologic settings, such as mountain belts \citep{hovius96}, fault
 blocks \citep{talling97}, submarine canyons \citep{inman76,orange94},
valleys of volcanic islands \citep{wentworth28,kopi86}, sapping canyons
\citep{hkh88,sbws95}, and Martian gullies \citep{malin00}.  The
ubiquity of the periodicity implies a common phenomenon but not a
common cause.  Indeed, as we now proceed to review, theoretical
formulations of erosion initiated by seepage and overland flow
indicate that characteristic spacings are expected, but not
necessarily for the same reasons.  It is unclear whether the various
forms of periodic channel spacing are due to a few basically similar
mechanisms or if there are many essentially different explanations.
There are many conceivable explanations for the periodicity.  We
approach this rich phenomenology by a detailed study of one specific
example that is particularly accessible to study.

\begin{figure}
\begin{center}
a)\includegraphics[width=8cm]{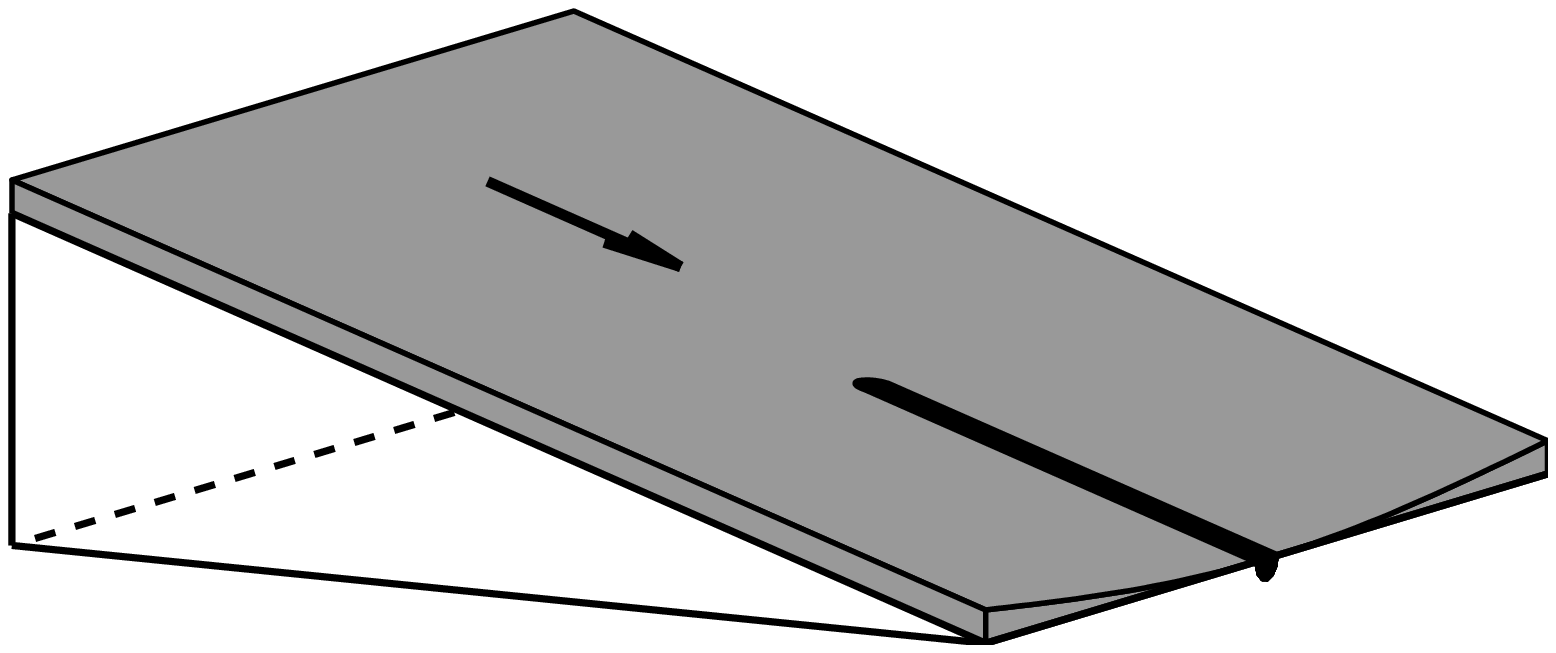}\\
b)\includegraphics[width=8cm]{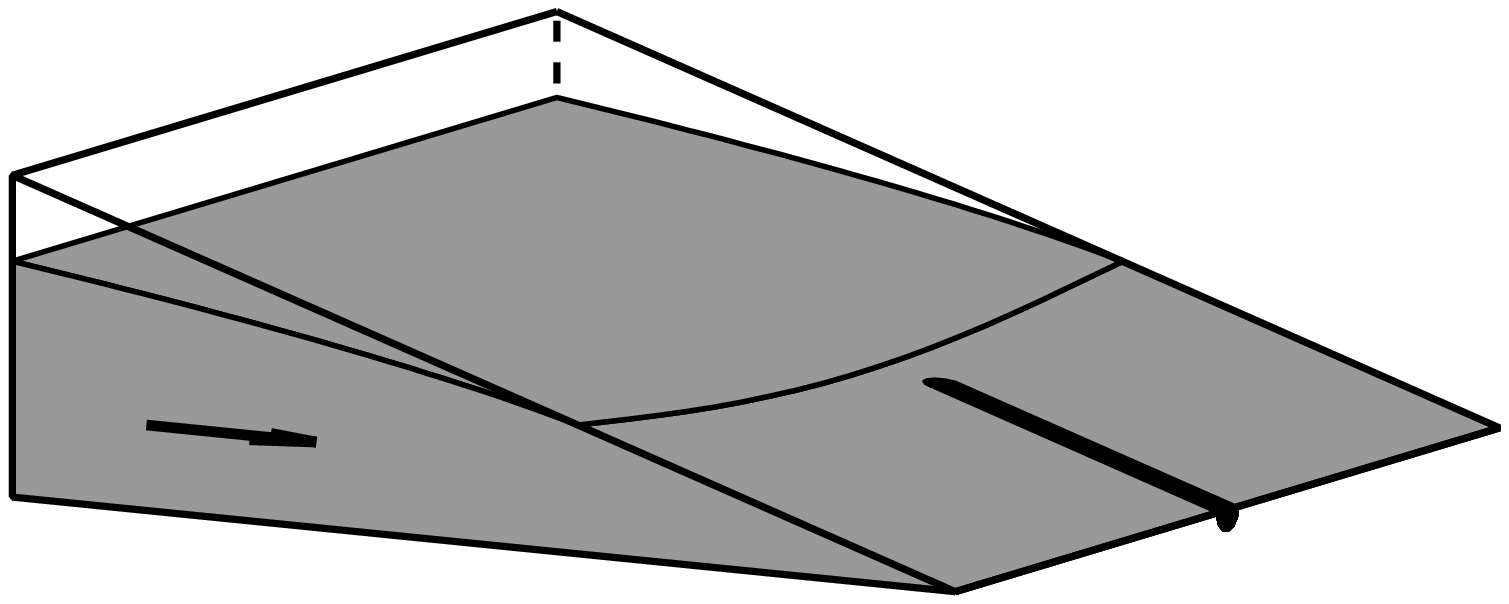}
\end{center}
\caption{Schematic view of two erosive processes: overland flow (a)
and seepage erosion (b).  In overland flow a thin sheet of water
(gray) flows over the surface, eroding it as a result of the shear
stresses induced on the surface by the flow.  In seepage erosion the
fluid flows through the porous subsurface and erodes surface material
after it emerges at the surface.  Both processes can produce channels
(the thick black lines).
\label{cartoons}}
\end{figure}

Broadly stated, two ``end-members'' delimit processes of
channelization \citep{kirkby67,dunne80}.  At one extreme is 
erosion due to {\em overland flow}, wherein shear stresses imposed by
a thin sheet of water flowing downhill act to erode a surface
\citep{horton45}.  At the other extreme is erosion due to {\em
groundwater flow} \citep{kirkby67,dunne80,dunne90}.  When a subsurface
flow erodes a porous material as it emerges from the material, the
process is called seepage erosion (or sapping). Overland flow and seepage erosion are
illustrated schematically in Figure~\ref{cartoons}.

A continuum model for overland flow was given in a landmark paper by
\cite{smith72}. 
It was shown that under certain circumstances the eroding surface is
unstable to sinusoidal channel-like patterns.  However they found that
the most unstable wavenumber was infinite, implying that the most
unstable wavelength is at the microscopic granular scale.  Noting this
deficiency, \cite{izpa95} provided an alternative theory based on two
components: a threshold condition for erosion and analysis of the full
equations for shallow overland flow.  The inclusion of the threshold
condition yields an instability at a finite wavelength, a finite
distance from the divide.  This result shows that periodic
channelization can occur entirely as a consequence of its own
dynamics.
Under different circumstances, a finite wavelength can even emerge
without any threshold condition \citep{izpa00}.

The theory of seepage erosion is less well developed.  As stated by
\cite{dunne90}, ``there have been no formal quantitative studies of
the relation between the hydrogeologic properties [of material
undergoing seepage erosion] and the average spacing of channels.''
\cite{dunne80,dunne90}, however, provides a physically appealing
description.  The pressure head in a horizontal or tilted plane leads
to high pressure gradients at places of high local curvature, and
hence presumably to enhanced erosion.  This instability leads to the
growth of channels.  A similar but more complex situation is depicted
in Figure~\ref{cartoons}b, where a channel on an inclined slope is
eroded below the point where the water table intersects the surface.
Once a channel forms, internal flow lines are focused toward it.  As
we show below, the focusing occurs over a large but finite transverse
extent, which is presumably related to a characteristic channel
spacing, the presence of which requires no heterogeneities in material
properties.

Documented experimental studies go back at least a century
\citep{jaggar08}.  In the 1980's, seepage experiments
\citep{khm85,kopi86,schumm87,howard88,ksp88} were frequently concerned
with analogy to Martian valleys.  The question of whether ancient martian
channels are created by seepage or vast amounts of surface water
remains open today \citep{baker82,lm85,bkblh90,aha02a}.  Our own
experiments are motivated by terrestrial analogs.
Previous experimental work suggests that groundwater piracy plays a
dominant role in the development of channels.  \cite{hm88} studied the
rate of seepage erosion in a narrow two-dimensional flow tank.
\cite{how94a,howard95} has carried out computer simulation modeling of
valley development by groundwater sapping.

The present study is a first report of our effort to conduct field
studies, experiments, and theory simultaneously to address the
formation of channels, and rhythmic channelization in particular, by
fluvial erosion when substantial groundwater flow is present.  The
following three sections cover, respectively, field study, experiment,
and theory.  The last section contains conclusions that we are able
to draw so far.

\section{Field Study: Beach Rills}

Beach rills \citep{komar76} can serve as inspiration for larger-scale
drainage channels forming over geologic time periods
\citep{higgins82}.  While incised channels are not uncommon on sand
beaches, periodic channelization is rarer.  Two suitable sites where
small incised channels form when the tide recedes are located on Cape
Cod.  One is at Provincetown, Massachusetts, a few minutes walk north
of the main pier, the other at Mayo Beach in Wellfleet, Massachusetts.
At high tide the sand is soaked with water, which seeps out as the
water recedes.  The tidal difference at both locations is typically 3
meters.  All channels are erased when the tide rises again and they
reform at different locations during the next tidal retreat.  One can
observe the generation of the channels on a daily basis.

\begin{figure}
\centerline{\includegraphics[width=13cm]{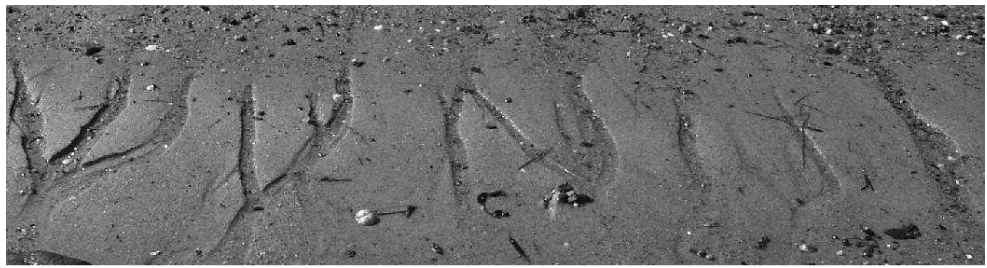}}
\caption{Beach rills observed in Provincetown, Massachusetts on 30
August 2001.  The shoreline is parallel to and below the picture.
The distance from left-to-right is approximately 2~m.}
\label{beachrills}
\end{figure}

The channels found on the beach are sharply incised
(Figure~\ref{beachrills}).  They have steep side walls so that their
cross section becomes almost rectangular and they are about one
centimeter wide.  The sand grains are fairly distributed in size, around a
mean of about 1mm.  At the Provincetown site, the average channel spacing,
based on over 200 measurements, is about 40cm.  
The spacing is measured near the head of the rills.
Depending on the
height of the tide, we observe 1-5 generations (bands) of channels,
with typical lengths of 0.7-2m.  Many channels begin at pebbles,
others start without visible perturbation.  Further downhill, channels
often merge and braid.  The beach slope is 10-12\%, terminated by a
1\% slope that emerges only at low tide.  At the Wellfleet site, the
spacing is about 1m, based on 38 channels, and the slope is 4\%, also
terminated by a lower slope.  There are no pebbles at this site.

We have observed channels several times during their initiation and
growth at the Provincetown site.  A vertical test hole allows us to
monitor the ground water level as a function of time.  The water table
inside the beach sinks 1cm every 3~minutes during the hours before
channel initiation.  During the same time, the sea level sinks about
three times as fast.  Hence, the height difference between water table
and sea level increases steadily until, about three hours after high
tide, the pressure head suffices to form channels.  

The channels clearly form outside the water and are not directly
related to the waves that hit the beach.  Waves arrive occasionally
and wash water up the beach, subsequently leading to a sheet of water
flowing downhill.  Pebbles interfere with this flow and temporarily
cause cone-like flow patterns.  These disturbances in the flow can
lead to an incised streak downhill of the pebble which eventually
grows into a channel.  The initial incision happens within seconds or
less and leaves a streak on the order of 10cm long.  Not all channels
are initiated this way.  Another frequent initiation mechanism results
from ``fingers'' of runoff water from any sources higher up (water
ponding behind stones or from higher-lying, previous channels).  Areas
without surface defects sometimes do not develop any channels.  At
other locations even tiny perturbations suffice for initiation and
empty areas are ``filled in'' with channels.  As far as one can see on
the surface, the periodicity is established promptly (within a minute
or so).

Once a channel has initiated it grows downhill.  Sand is washed away
at the lower end of the channel by water running in the channel.  At
the downward end of the channel, where the longitudinal extension
occurs, the water appears to be fully loaded with sediment.  There is
mass wasting on the sides so that the channels widen slowly; they also
deepen slowly.  Sand grains are transported by the water flowing in
the channel.  The rills grow to their final size within minutes to tens of
minutes.

\begin{figure}
\centerline{\includegraphics[width=6cm]{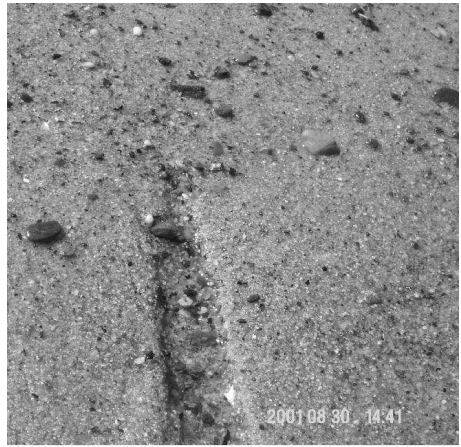}}
\caption{An incised channel on the beach filled with running water.
The image is about 15cm wide.  A pebble is located at the channel
head.  The water in the channel emerges from underground.
\label{choff}}
\end{figure}

Figure~\ref{choff} shows a channel filled with water.  The channel
could be fed either by surface water that seeped out above the
channel, or directly from groundwater provided from the bottom and
sides of the channel.  We found that they are directly fed by
groundwater, for the following reasons.  The film of water above the
channel shows no movement visible to the eye.  Small grains, placed by
us on the surface, diffuse only slowly at speeds that are insufficient
to provide the large amounts of water seen inside the channels.  The
channels carry water for quite some time after the initiation and long
channels continue to be filled with water when the surface at its head
is already dry.  All this shows that the channels are fed directly by
groundwater.

The beach rills provide a natural example of rhythmic channelization
and the observations suggest that they require seepage for their
explanation.

\section{Seepage Experiments}
\subsection{Experimental Setup and Channels}

\begin{figure}
\begin{center}
a)\includegraphics[width=12cm]{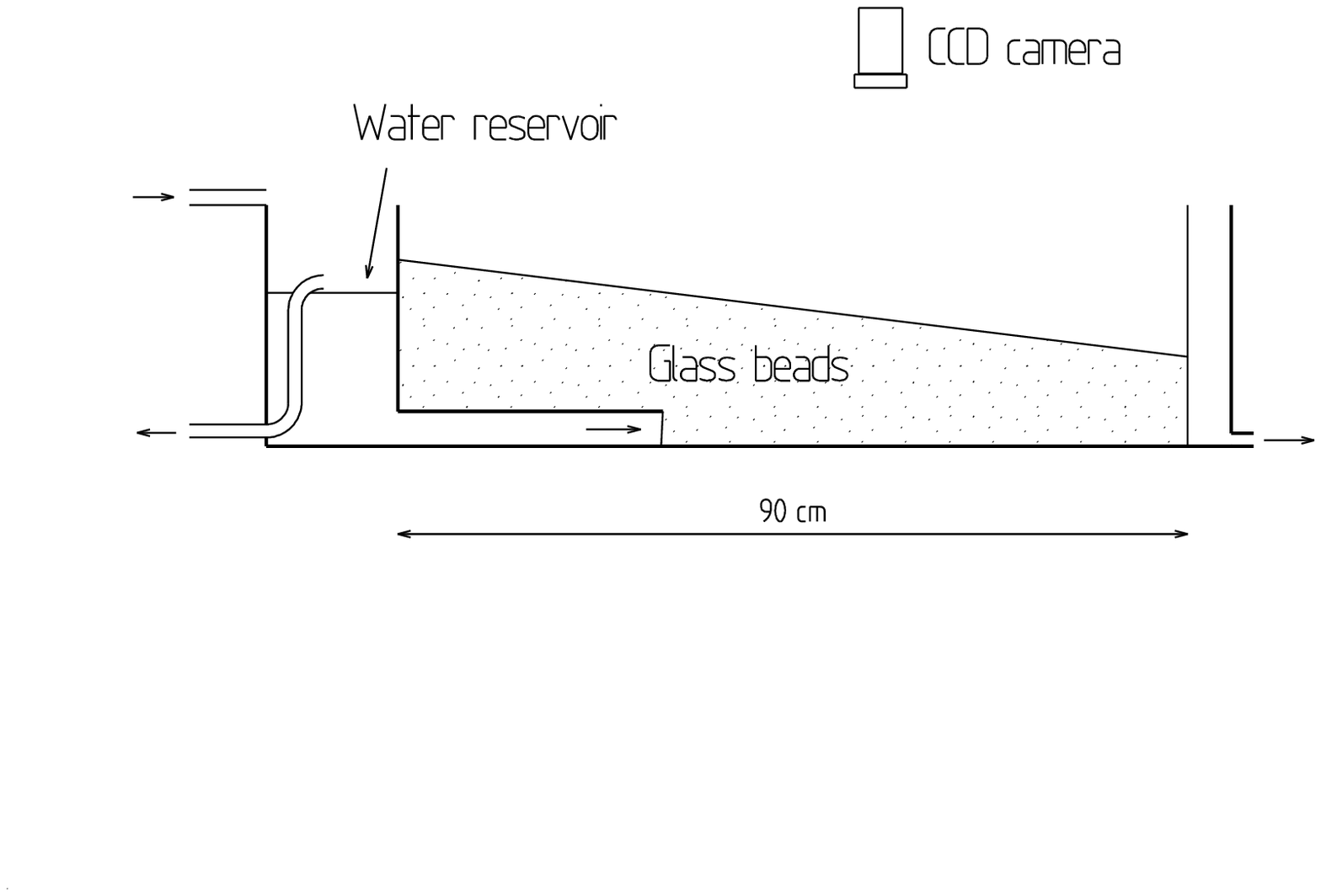}\\
b)\includegraphics[width=7.5cm]{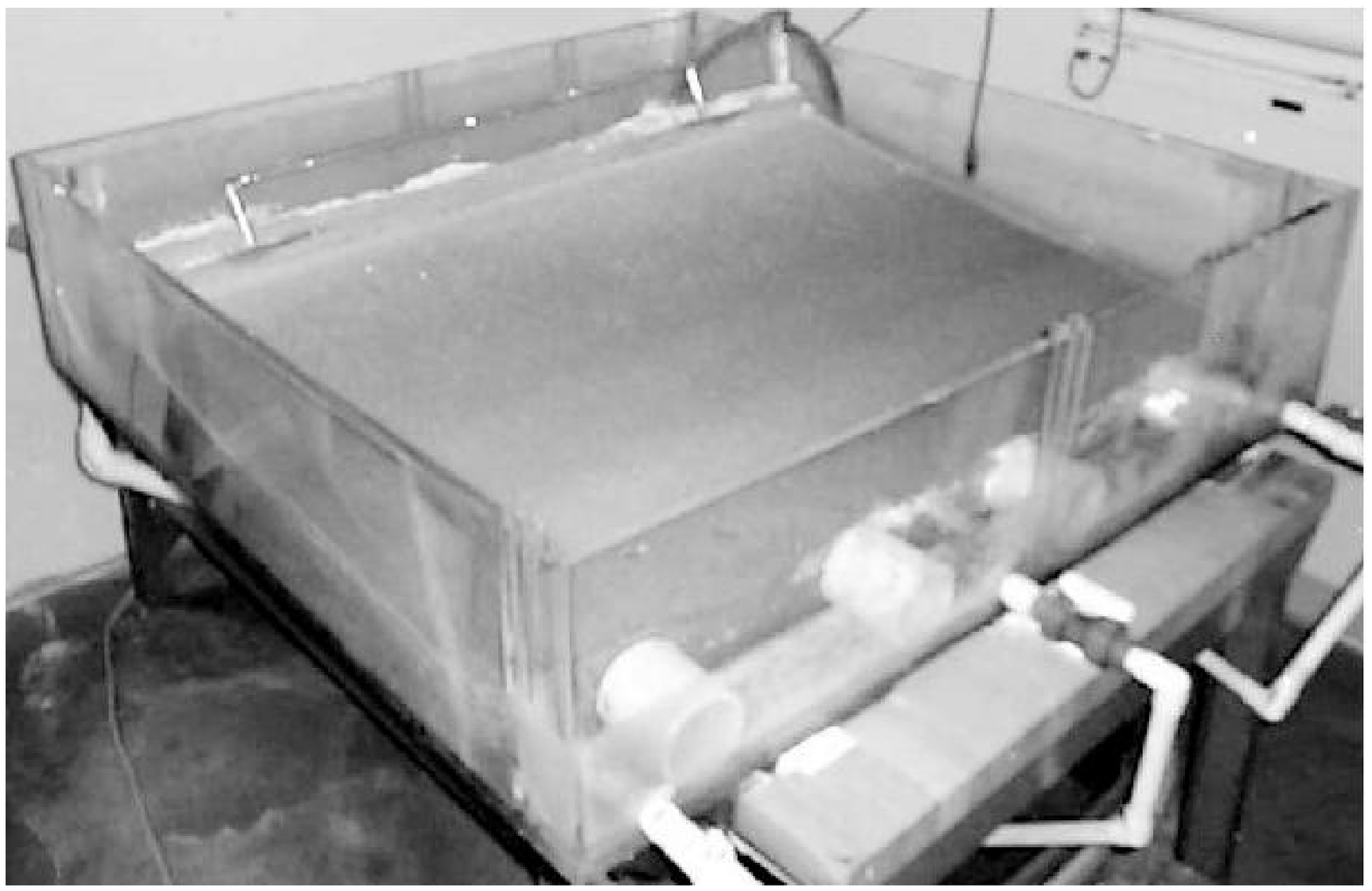}
\end{center}
\caption{The experimental setup. a) A two-dimensional schematic cross
section. b) The actual apparatus.}
\label{exp-setup}
\end{figure}

Prior experimental studies of seepage erosion
\citep{khm85,kopi86,howard88,ksp88} have observed a few periodic
channels.  Our own experimental setup is shown in
Figure~\ref{exp-setup}.  The plexiglass box is 120cm wide and
effectively 90cm long.  It is filled with 0.5mm glass beads.  Water is
pooled in the back and kept at a constant level using an overflow
tube.  The water seeps through the granular medium and creates
channels on the slope, observed by a CCD camera.  
Sideways
illumination makes the channels clearly visible in the image.  The
experiments are carried out with a steady water reservoir, unlike the
beach where the water table is receding.

A ledge protrudes at the bottom to increase the flux through the system.
A minimum flux is required to dislodge a grain on an inclined surface
which is below its angle of repose.  The ledge makes it possible to
achieve the threshold in a finite sized experimental apparatus, by
allowing us to increase the height of the water column in the reservoir
above the height of the granular bed. The resulting flux is also similar
to that observed at the beach when channels are formed. Thus the ledge
allows us to reproduce the processes observed on the beach in a finite sized
experimental system.  We did not observe any channels without the ledge.
                          
The capillary height in the medium is measured to be 25mm. The
permeability is determined from flow through a U-tube.  A water column
of height $\Delta h$ seeping through a porous medium over a distance
$L$ flows with a velocity of $v=k \Delta h/L$.  The seepage
coefficient $k$ is 3~mm/s.  Sand from the beach, with its moderately
larger grain size, has a moderately higher permeability,
$k\sim$8~mm/s.

\begin{figure}
\begin{center}
\begin{tabular}{ll}
a) h$_W$=13.8cm, s=11\%, t=28 min &
b) h$_W$=16.2cm, s=11\%, t=18 min \\
\includegraphics[width=66mm]{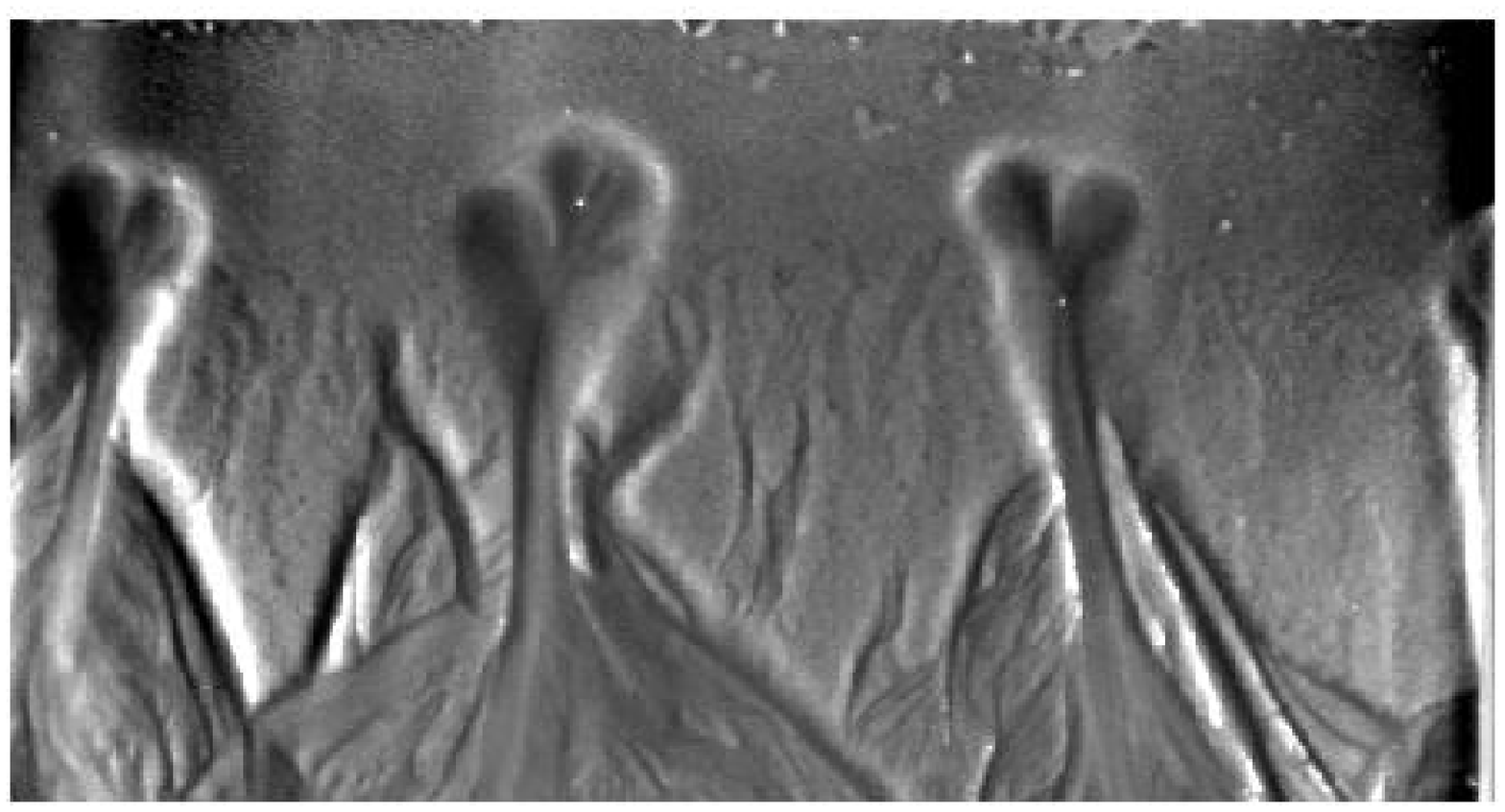}
&\includegraphics[width=66mm]{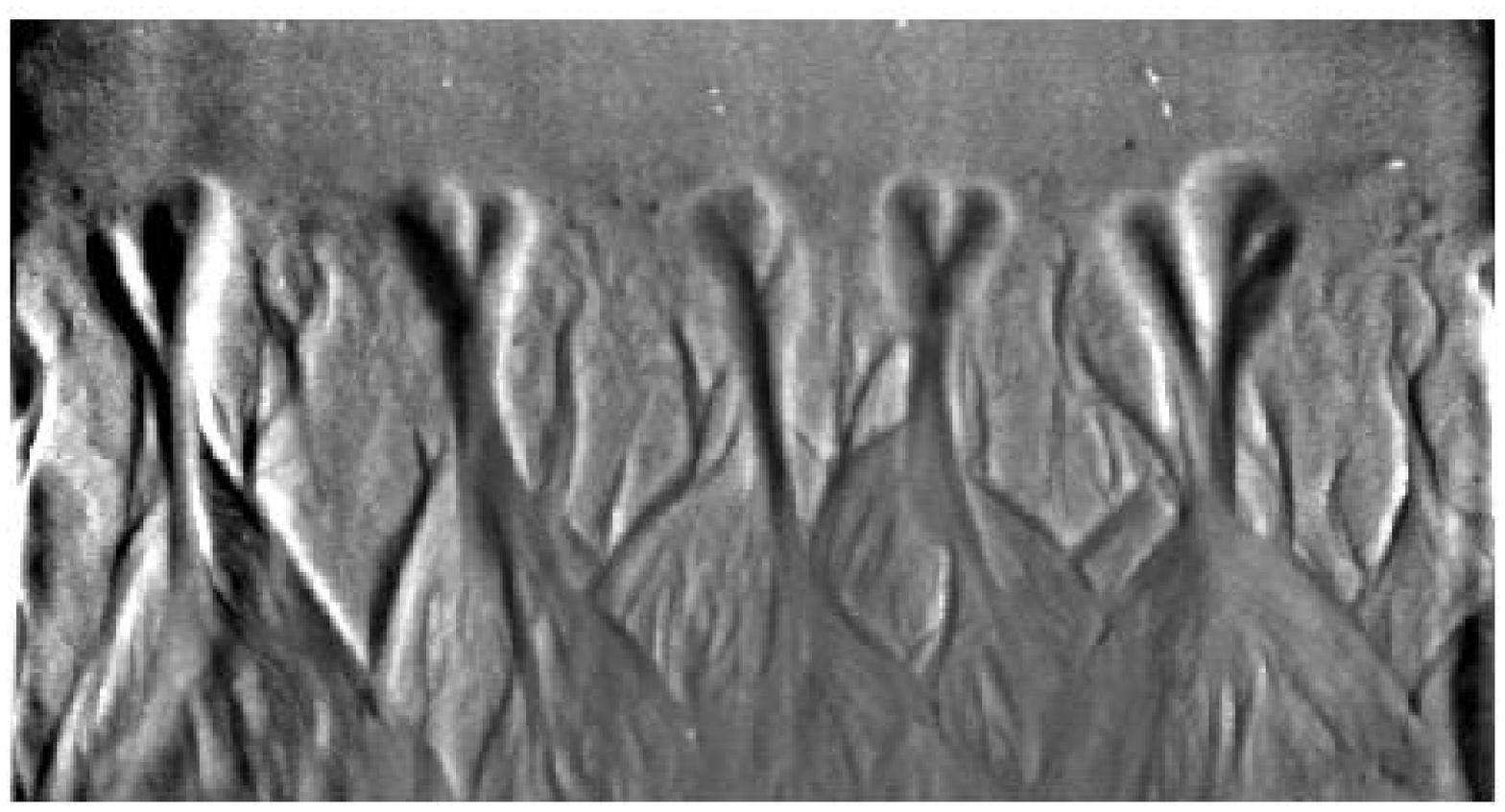}
\\
c) h$_W$=15.8cm, s=10\%, t=18 min &
d) h$_W$=13.4cm, s=16\%, t=2 min\\
\includegraphics[width=66mm]{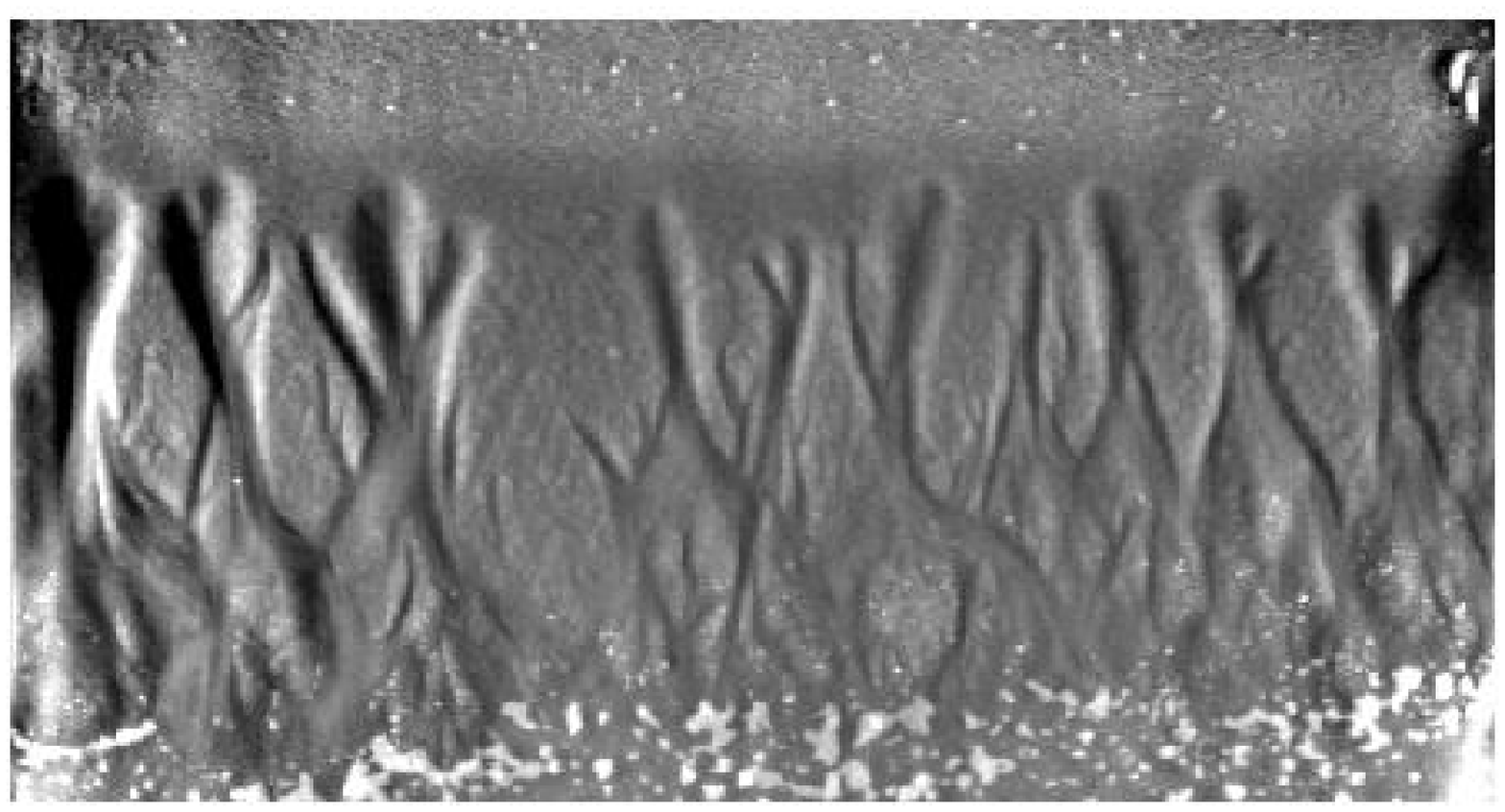}
&\includegraphics[width=66mm]{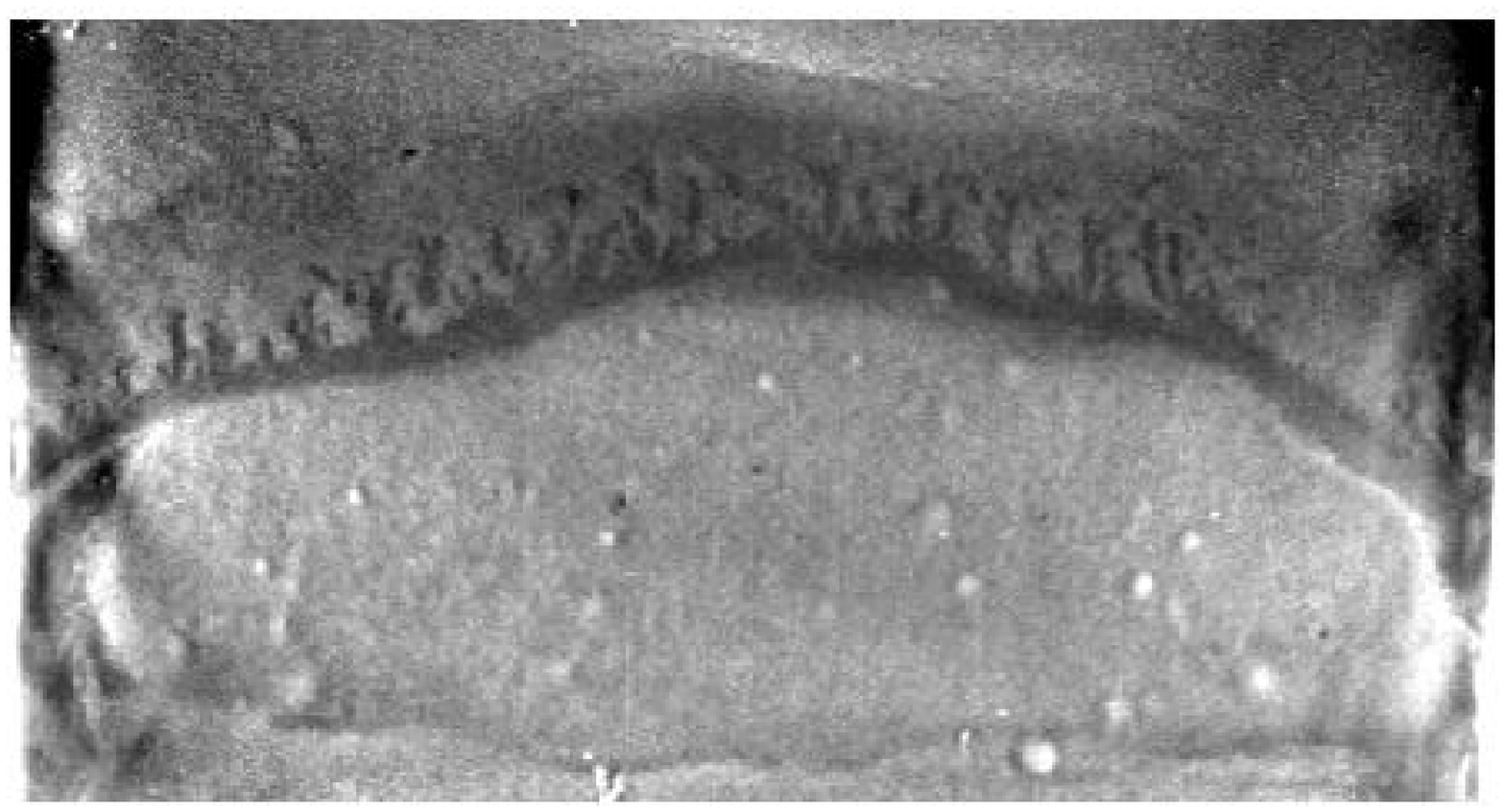}
\end{tabular}
\end{center}
\caption{The outcome of four laboratory experiments done at various
slopes $s$ and water column heights $h_W$.  The water source is toward
the top side of each image.  Illumination is from both sides.  The
image width is 120cm.  (a-c) Examples of multiple drainage
channels. (d) Example of a slump observed at high slope.}
\label{expchannel}
\end{figure}

Before the beginning of each experiment a planar, sloped surface is
created, then the water column in the back is filled to a prescribed
height $h_W$, and the camera is turned on.
The height $h_W$ of the water column is measured from the bottom of the tank. 
Figure~\ref{expchannel} shows
examples of laboratory channels. These images are taken less than half an hour
into the experiment. Figure~\ref{expchannel}c shows about 10 drainage channels.  
Figure~\ref{expchannel}d corresponds to a slump observed at high slopes.
In this paper, we only report
on patterns of fairly straight channels.  Other erosion patterns, some
of which are found by \cite{daerr02} in a different experimental
setup, are observed in various parameter regions.

\begin{figure}
\begin{center}
\includegraphics[width=8cm]{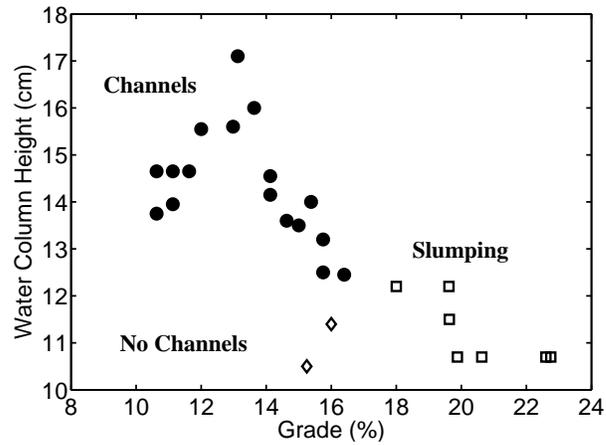}
\end{center}
\caption{Observed behavior as a function of slope $s$ and water column
height $h_W$.  Filled circles ($\bullet$) correspond to spontaneous
channelization.  High slopes lead to slumping of the material
($\square$).  No channels ($\Diamond$) form at low pressures and
slopes.}
\label{phase}
\end{figure}

Channels such as those in Figure~\ref{expchannel}a-c are observed within
a limited range of slopes and water column heights.
Figure~\ref{phase} summarizes the parameter regions for which
experiments were carried out.
In our early experiments, parameters were not accurately controlled,
but Figure~\ref{phase} corresponds to a later set of experiments
conducted with a constant experimental setup.
Channel formation is observed for slopes $\lesssim$17\%. This slope is
much less than the angle of repose of completely dry or saturated
glass beads which is approximately 43\%.  For the particular level of box
filling, channels are not observed below a water column height of
12cm.

The mere observation of regularly spaced channels in the experiment
immediately constrains possible explanations.  The medium is composed
of mono-disperse, cohesionless glass beads.  The fluid is clean water.
This is all that is required to create periodic channels. Reproducing
channelization in a clean and controlled environment is a major step
in studying the phenomenon and provides the basis for theoretical
modeling.

\subsection{Temporal Evolution}

\begin{figure}
\centerline{\includegraphics[width=13.5cm]{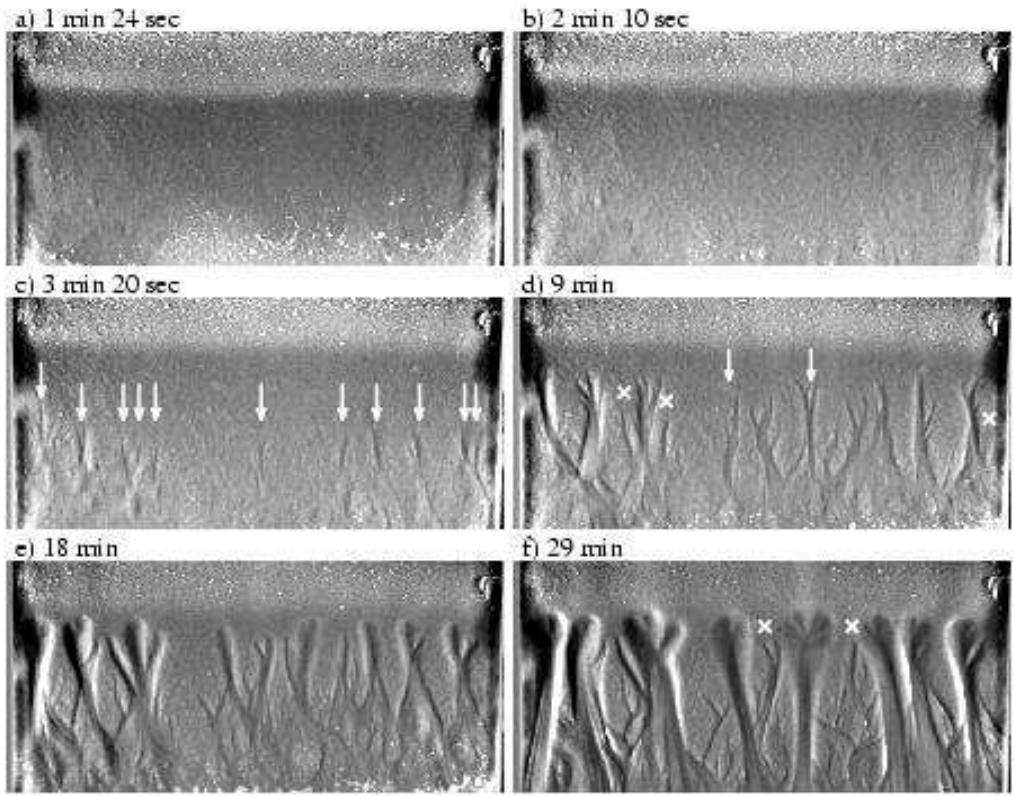}}
\caption{Periodic channels observed in the experiment.  The water
source is toward the top side of each image.  a) An initial sheet of
water, b) still no visible channels, c) incision of channels (marked
with arrows), d) channels grow, some additional ones have formed, and
others have dried up, e) continued growth, and f) the channels after
about half an hour.  New channels, not visible in the previous image,
are marked with arrows.  Channels that have discontinued growth
are marked with crosses.}
\label{expchannels}
\end{figure}

Figure~\ref{expchannels} shows a time sequence of six images of the
same experimental run.  The channels are initiated after a sheet of
water flows down the surface.  The sheet disappears and the channels
deepen, widen, and grow lengthward.  This is similar to the beach
rills, where channels are also initiated by surface water, then grow
downhill by washing away sand at the lower end of the channel.  Unlike
the beach, the experimental channels, in addition to growing downward,
also grow headwards until they stop at the point where the ledge ends.
Channels grow initially (quickly) downwards and then slowly upwards (headward).
After the channels have reached their maximum length they still widen
and occasionally meander and fan at the lower end.


The spacing of initial channels, Figure~\ref{expchannels}c, is also apparently
not random.  It may result from an instability in the initial sheet
flow \citep[e.g.][]{huppert82} or from the interaction of the sheet
flow with the eroding surface.  Any threshold for incision is likely
related to the sheet flow, rather than to the force exerted on the
grains by water emerging directly from underground.  

\begin{figure}
\centerline{\includegraphics[width=10cm]{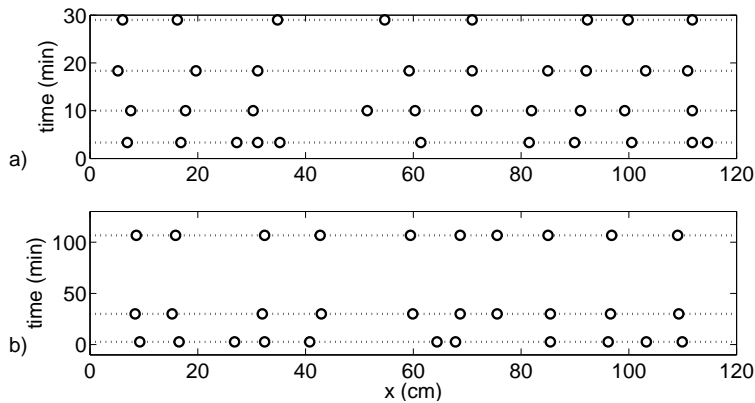}}
\caption{The position of major channels in two experimental runs as a
function of time.  Only some of the initial channels grow into large
channels and coarsening occurs with time.  The final spacing is more
periodic than the initial state.  Part (a) corresponds to the
experiment shown in Figure~\ref{expchannels}.}
\label{coars}
\end{figure}

Figures \ref{expchannels} and \ref{coars} show spatial coarsening over
the period of channel development.  Only some of the initial channels
grow into larger channels, while others cease their growth.  For
example, in Figure~\ref{expchannels}f there are several smaller channels
visible that have stopped growing.  This is especially the case when
another channel is nearby. Growing channels tap into the groundwater
supply of their neighbors, which then cease to grow due to the lack of
groundwater.  On the other hand, new channels still form after the
initial incisions, especially when far from existing channels.
Figure~\ref{expchannels}d shows two examples.  Both of these processes
-- the competition of nearby channels and the formation of new
channels away from existing ones -- eventually lead to a more regular
spacing.

In experiment (a) of Figure~\ref{coars} there are 11 initial channels,
of which 5 eventually cease to grow.  The other 6 initial channels
grow into major final channels along with 2 channels that formed at a
later stage.  In experiment (b) there are 11 initial channels, of
which 3 cease to grow and two additional channels form at a later
time.  These numbers indicate that many of the big final channels have
grown directly from small initial channels.  However, the selection of
which channels dry up and where new channels form is influenced by the
distance to neighboring channels, so that the spacing is dynamically
created or at least dynamically grown from the initial state.  Channel
locations have more of an imprint of the initial state than does the
spacing between them.  Incision behavior {\it and} groundwater piracy
contribute to the final state.  With more separation in time and space
between initial and final configuration one would expect an even
larger spacing, which has little to do with the initial state.

\section{Theory and Numerical Simulation}
\subsection{Governing Equation}

As a result of the observations on the beach and in the experiment,
the problem can be formulated theoretically in terms of flow through a
porous medium.  After channel initiation, the supply of water to the
channel is governed by groundwater flow.  The erosion at the channel
boundaries, especially at the lower end of the channel, is governed by
the amount of groundwater collected in the channel.  Our aim is to
determine the extent of the groundwater piracy which limits the growth
of channels.

The presence of long tubular structures in the soil, or ``macropores,''
could significantly influence the flux of water through the medium.
We have not observed any such pores.  The material is cohesionless and
cohesionless grains cannot sustain voids and therefore macropores are
not expected.  According to the observational and theoretical
evidence, we discard this possibility (no gophers). 
Local density variations may be possible.

In a porous medium, the velocity is given by Darcy's law \citep{kochina62}
\begin{equation}
\vec v=-k \left({\vec \nabla p\over\rho g}-\hat g\right),
\label{eq:darcy}
\end{equation} 
where $k$ is the seepage constant, $p$ is the
pressure, $\rho$ the density, $g$ Earth's acceleration
constant, and $\hat g$ the unit vector in the direction of gravity.  
(The permeability $\kappa$ of the medium is directly related to the seepage coefficient 
through $\kappa/\mu = k/(\rho g)$, where $\mu$ is the viscosity of the fluid.)
Since the fluid is incompressible, $\vec\nabla\cdot \vec
v=0$, and it follows from~(\ref{eq:darcy}) that
\begin{equation}
\nabla^2p=0.  
\label{eq:laplace}
\end{equation}

\begin{figure}
\centerline{\includegraphics[width=13cm]{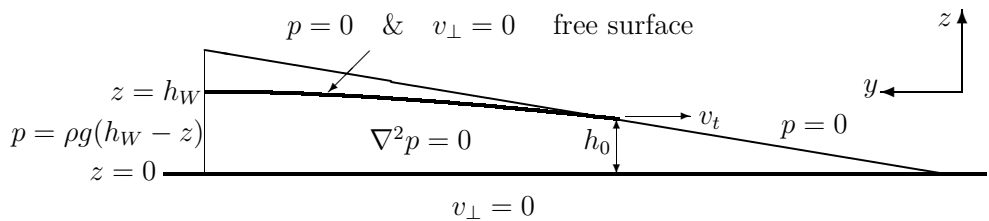}}
\caption{Boundary conditions for a slope on top of an impermeable
layer.  
The velocity component normal to the surface is denoted by $v_\perp$.
For the beach, the bottom boundary must be replaced by
tailwater.  The figure also defines $h_0$, the height of the
intersection between water table and surface, and $v_t$, the absolute value of the velocity
at this intersection.}
\label{bc}
\end{figure}

Boundary conditions are shown in Figure~\ref{bc} for a stationary
situation.  
The bottom of the tank is impermeable, the water column in the back is
represented by a linearly varying pressure, and atmospheric pressure
is applied at the seepage surface. 
At the water table there is atmospheric pressure, $p=0$,
and the velocity is tangential.  The shape of this free surface
adjusts such that both boundary conditions are satisfied
simultaneously.  The pressure distribution and the shape of the free
surface are independent of the parameter $k$, since neither the
Laplace equation (\ref{eq:laplace}) nor the boundary conditions depend
on $k$.  The shape of the water table is also independent of
$g$, since the boundary conditions and the Laplace equation
can be written in terms of $p/(\rho g)$.  (The limit $g\to 0$ poses no
physical problem, since for $g=0$ every shape is permissible.)  While
velocities depend on $k$ and $g$, the water table is purely
determined by the geometry of the boundaries. 
This is also obeyed by the known exact solutions for unconfined flow
\citep[][Chap. VII]{kochina62}.

In the above formulation of flow in a porous medium, capillarity and
evaporation are ignored.  Evaporation should be negligible over the
time period the channels form.  On the beach and in the experiment the
capillary height exceeds the depth of the channels, and hence the
capillary forces may prevent the detachment of the water table from
the surface.  However, the capillary effect vanishes for larger scale
landscapes.  
Further consideration of the potential role of capillary forces will
be necessary before detailed quantitative comparisons with experiments
are made.
We expect however, that our qualitative conclusions will
be independent of these issues, because ground water piracy takes
place in either case.

\subsection{Numerical Method; Planar Sloped Surface}

\cite{kochina62} provides a comprehensive collection of analytic
solutions to problems of groundwater movement, but most geometries
need to be solved numerically.  We solve the Laplace equation
numerically with a conventional relaxation method on a regular grid.
The scheme uses the pressure at the six neighboring grid points and
iterates until the pressure field has converged.
At the free boundary, cells are no longer treated as
rectangular, but in the three-dimensional finite difference
approximation of the Laplace operator, the exact location of boundary
point(s) is taken into account.  At the free boundary, atmospheric
pressure is imposed and the free surface is moved until it is stationary,
using a simple first-order time integration.  The time-advanced grid
points representing the free surface no longer lie on a regularly
spaced grid and are interpolated back onto a rectangular grid
at each integration step.  In three dimensions, when the free surface
is two-dimensional, the interpolation fits a plane through three
nearby points.  The three points are chosen such that without
variations along the $x$-axis, one exactly recovers a one-dimensional,
linear interpolation scheme.  Hence, three-dimensional simulations
without variation along the $x$-direction yield exactly the same
results as two-dimensional simulations.  The numerics are tested by
comparison with the analytic solution of the vertical dam 
\citep[][eqs. VII.10.34, 10.35, 10.41]{kochina62}
and with analytical estimates for the flux through a wedge
\citep[][eq. VII.12.12]{kochina62}.


\begin{figure*}
\includegraphics[width=13.5cm]{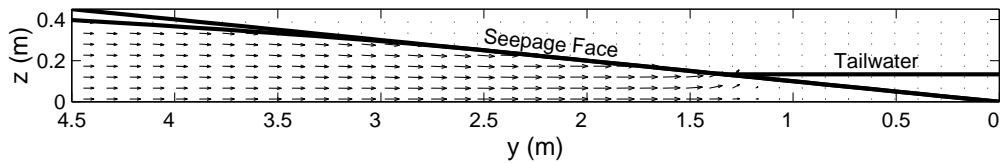}
\caption{Cross section of the beach with water table and velocities
obtained by numerical solution of the porous media equations.  At high
tide the sand is soaked with water.  As the tide recedes, water seeps
out along the beach face, starting at the point where the water table
intersects the surface.  The effect of capillarity is neglected.}
\label{beach}
\end{figure*}

Figure~\ref{beach} shows the velocity field in a geometry similar to
the beach (without channels).  As the height of the water table
decreases along the horizontal direction, the velocity increases as a
result of the incompressibility of the flow.

There are approximate solutions for a wedge without tailwater
\citep{kochina62}.  If the slope $s$ is small and the wedge is long,
then the height $h_0$ at which the water table intersects the surface
(see Figure~\ref{bc}) is approximately $h_0\approx (Q/k) (1/2+1/s)$,
where $Q$ is the flux per unit width and $k$ the seepage constant.
The average horizontal velocity at the point where the water table
intersects the surface is $\bar v = Q/h_0$, hence
\begin{equation}
\bar v \approx {2ks \over 2+s}\approx ks.
\label{eq:vt}
\end{equation}
The velocity increases with slope, because $h_0$ decreases with
slope.  Since the velocity changes little along the vertical (see
Figure~\ref{beach}), the average velocity $\bar v$ is a good
approximation to the velocity at the intersection,
$v_t\approx \bar v$.  Hence, the velocity with which water emerges at
the seepage face is higher for higher slopes.  Our numerical
simulations show that for slopes of 10\% or smaller the presence of
tailwater does not change the velocity $v_t$ significantly.

\subsection{Wentworth Instability}
\label{subsec:wentworth}

So far we have only considered seepage without channels.  Channels can
be represented in the simulation by applying the atmospheric boundary
condition at a properly shaped surface.  This shape is the planar
seepage surface with channels carved out.  The depth of water running
within channels is neglected.

\begin{figure}
\begin{center}
a)\includegraphics[width=9cm]{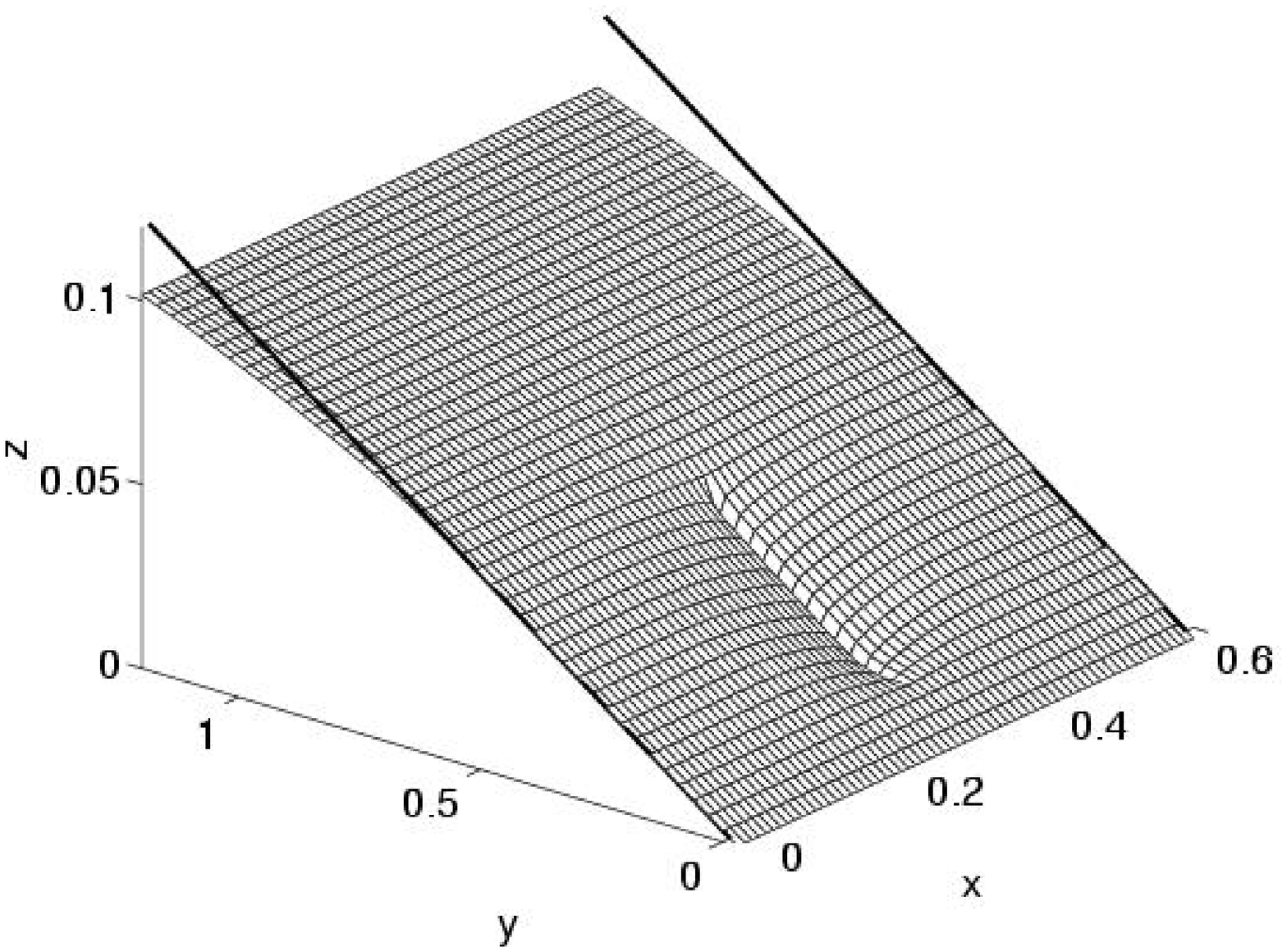}\\
b)\includegraphics[width=9cm]{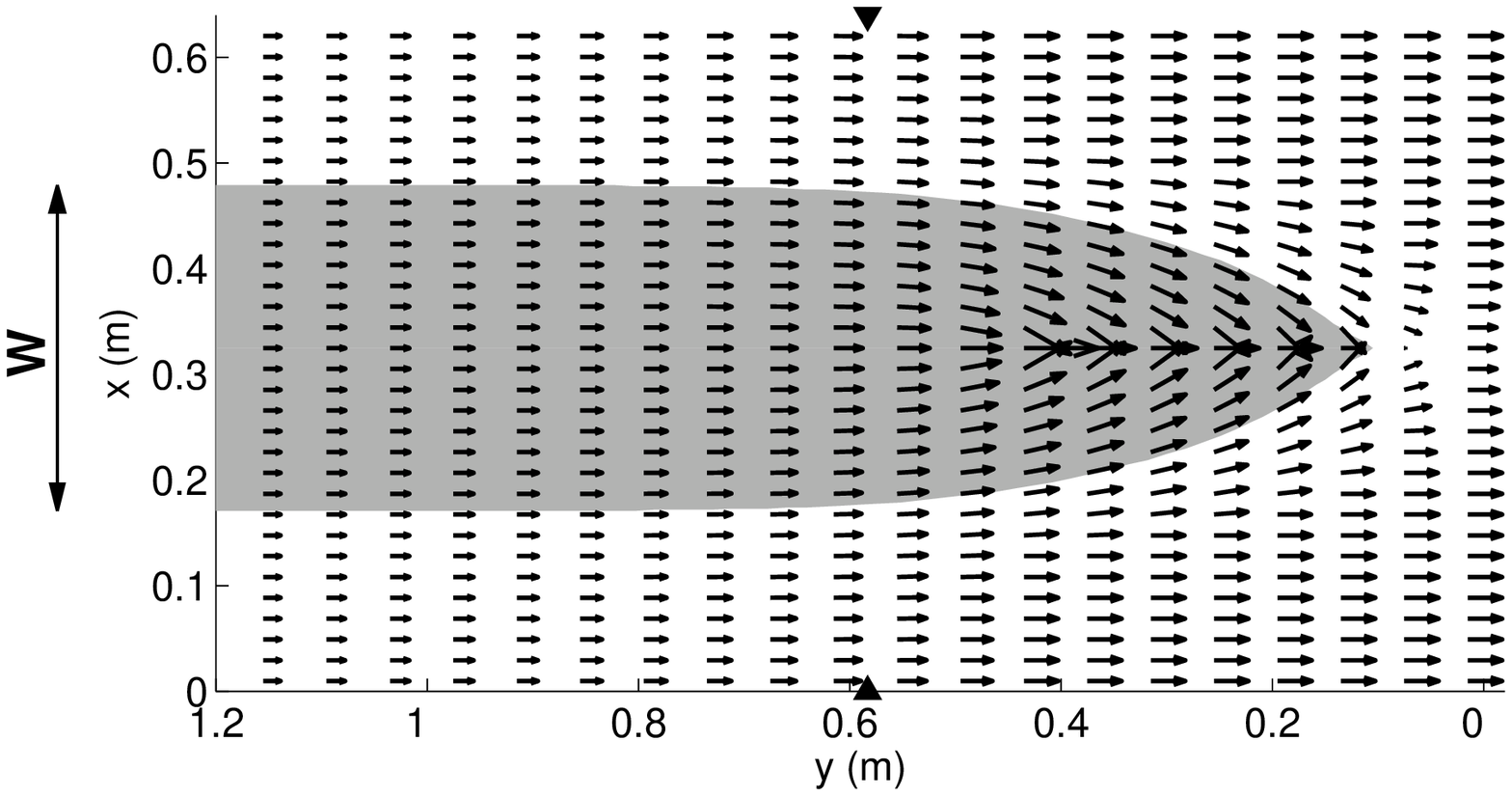}
\end{center}
\caption{a) Deformation of the water table due to an outlet channel,
as obtained by numerical solution of the porous media equations.  The
aspect ratio in (a) is distorted to emphasize vertical changes.  The
actual slope of the wedge is 10\%.  The deformations of the water
table increase the flux of water into the channel.  b) Velocities on
the surface of the water table, showing the influence of an outlet
channel on the flow field.  (The resolution of the calculation is
higher than the number of arrows plotted.)  Water within the shaded
region flows into the channel.  Black triangles mark the intersection
of the water table with the surface.
\label{wentworth+birdview}}
\end{figure}

The presence
of a channel perturbs the flow field.  Once a channel is formed, flow
lines focus toward it, but furthermore it leads to inclination of the
underground water table in response to an outlet channel.
Figure~\ref{wentworth+birdview}a illustrates the deformation of the
underground water table.  This brings even more water to the channel
and promptly acts as an efficient instability mechanism.  This
scenario was qualitatively described by \cite{wentworth28} in the context of
Hawaiian valleys.  Hence, we refer to it as ``Wentworth Instability''.
Figure~\ref{wentworth+birdview}b shows the velocity field on the water
surface.  The channel collects flow over a certain lateral distance,
hence suppressing the growth of other channels nearby, due to
groundwater competition.

The area over which the channel draws water can be visualized by
following the flow lines which reach from the water reservoir to the
bottom end of the channel.  The vertical component of velocity, $v_z$,
is small and approximate flow lines are obtained by integrating the
horizontal components of the velocity field.
The geometry of the region of influence (the shaded region in
Figure~\ref{wentworth+birdview}b) is again purely determined by the
geometry of the boundaries, including the dimensions and position of
the channel.  It is independent of density, permeability, and other
material properties.  The extent of this drainage area (or drainage volume) describes the screening
effect of the channel and is therefore a relevant mathematical problem
in this context.

\begin{figure}
\centerline{\includegraphics[width=9cm]{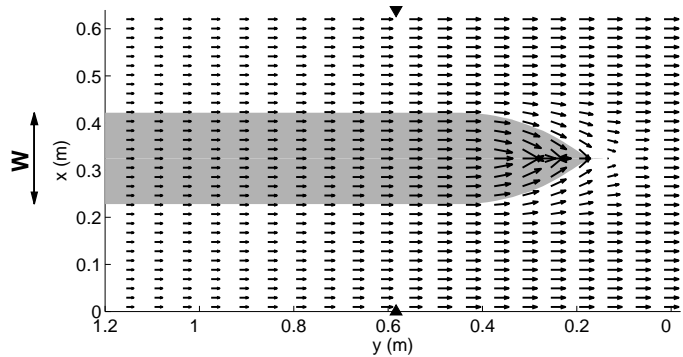}}
\caption{Compared with Figure~\ref{wentworth+birdview}b, the region of
influence changes with the length of the channel.  All other
parameters are the same as for Figure~\ref{wentworth+birdview}.}
\label{conewidth}
\end{figure}

The overall downhill
velocity $v_y \approx v_t$ is approximately constant along the
seepage face.  The width $\mathcal{W}$ over which water is collected
into a channel of length $\ell$ approximately obeys $\mathcal{W}/\ell
\propto v_x /v_y$.  As discussed above, variations of $v_x$ and $v_y$
in space are slow, and hence the width $\mathcal{W}$ is roughly
proportional to channel length $\ell$.  The constant of
proportionality depends on channel depth (via $v_x$) and on the slope
(via $v_y$), unless $\mathcal{W}$ extends to a distance $O(\ell)$
over which $v_x$ changes.  Figure~\ref{conewidth}, in comparison with
Figure~\ref{wentworth+birdview}b, shows this effect in full numerical
simulations.  The width of area drawn into the channel becomes smaller
and smaller for shorter and shorter channels.

Groundwater piracy should relate to channel spacing.  Since the
lateral claim of channels grows proportionally with their length, it
is small initially.  Small channels have only a small influence on the
flow field.  Long channels on the other hand will compete with nearby
channels.
With the growth of the channel, its region of influence on the
groundwater movement grows.

\subsection{Discussion}
\subsubsection{Length-scale for the width of the subsurface drainage area}

In this section we further discuss the width over which channels draw
groundwater.

\begin{figure}
\centerline{\includegraphics[width=13cm]{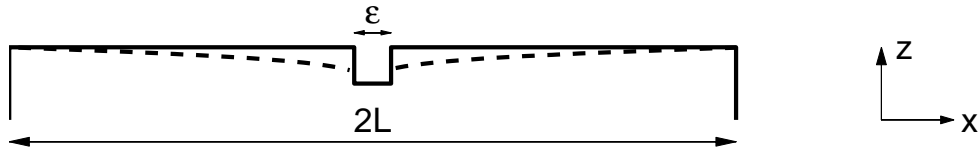}}
\caption{Deformations of the water table (dashed line) in response to
a channel of width and depth $\epsilon$.  Note the large horizontal
extent, $O(L)$, of the deformation.}
\label{scale_ill}
\end{figure}

Figure~\ref{scale_ill} illustrates, in a two-dimensional auxiliary
problem, that the influence of a channel of size $\epsilon$ is not
$O(\epsilon)$, as it would be without a free surface, but $O(L)$,
where $L$ is the distance to the boundary.  Without free surface, if
only the atmospheric pressure boundary condition were applied, the
influence of the channel would only reach $O(\epsilon)$.  (Since the
flow is driven by height differences, the length-scale of the water
table deformation translates into the length-scale over which the
velocity component, $v_x$, changes.)  The velocities
themselves strongly depend on $\epsilon$, but the scale over which
they change does not.  As a consequence, the lateral influence of the
channel can be much larger than its width or depth.  The {\it
deformations of the water table}, although small, {\it enhance the
lateral influence of the channel} and lead to quick growth of the
channel.

This length-scale can be argued for as follows.  Due to
incompressibility $v_x(x) h(x) \approx \mbox{constant}$,
where $h(x)$ is the height of the free surface.
When the behavior is close to hydrostatic, $p(x,z) \approx \rho g
(h(x)-z)$.  The actual depth over which horizontal motion takes place
does not need to be known, but is assumed roughly constant with $x$.
Using equation~(\ref{eq:darcy}), the velocity is
\[
v_x(x,z)= -{k\over\rho g} {\partial p \over \partial x} \approx - k {dh(x)\over dx}.
\] 
Therefore, $ h(x) dh(x)/dx \approx \mbox{constant}$.  After
integration, the unknown constants can be expressed in terms of
$h(0)$, the uppermost point, and $h(L)$, the lowest point of the free
surface.  We obtain
\[
h^2(x) = h^2(0)-\left[h^2(0)-h^2(L)\right]{x\over L}.
\qquad\mbox{(approximation)}
\]
$h(L)$ cannot be determined within this approximation.  After
differentiation at $x=0$,
\[
h'(0)=-{h(0)+h(L)\over 2h(0)}{h(0)-h(L)\over L}
=O\left({h(0)-h(L)\over L}\right).
\]
The inclination at the boundary is comparable to the average slope of
the water table.  Hence, the water table is significantly sloped over
the entire length, not just near the channel.

\begin{figure}
\begin{center}
a)\includegraphics[width=6cm]{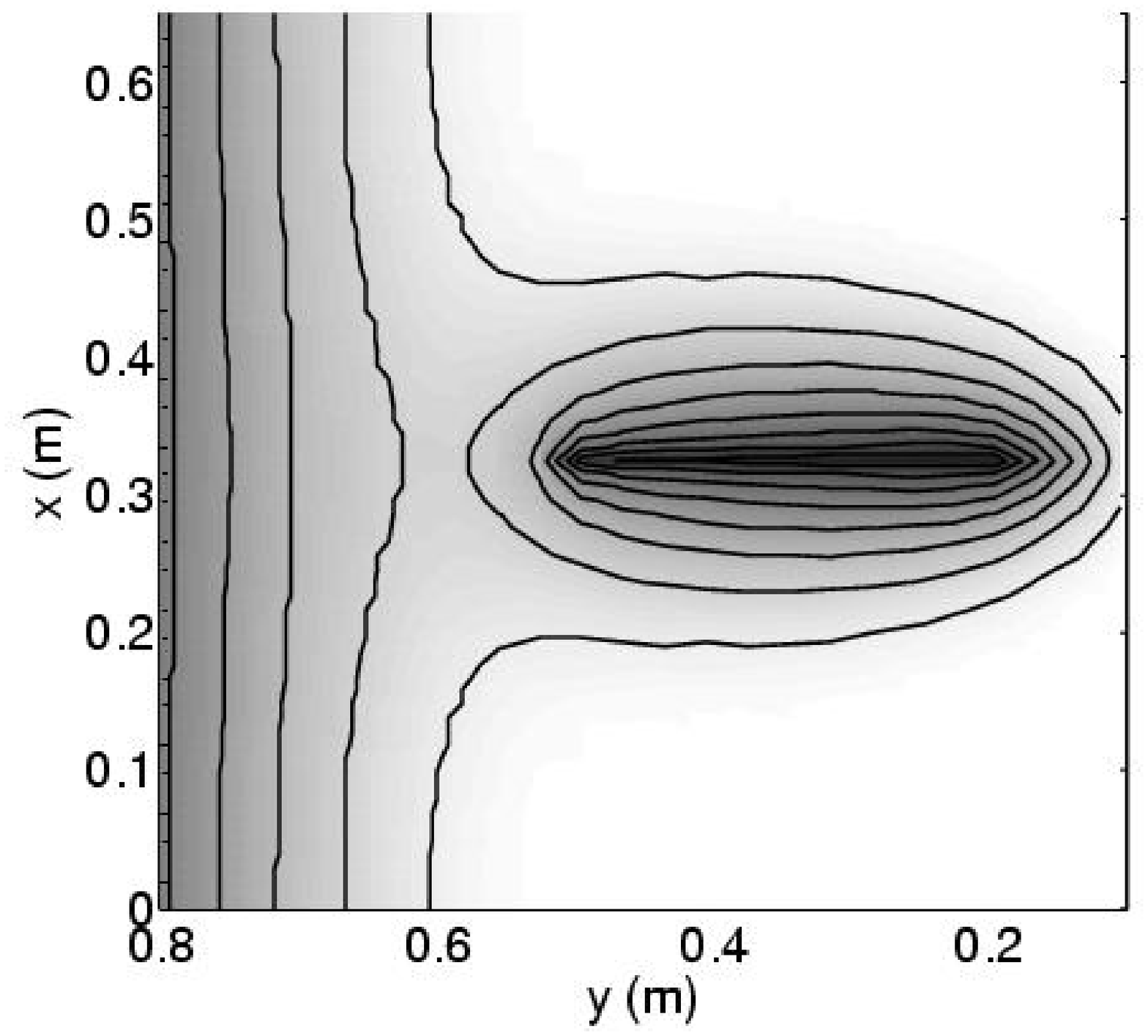}
b)\includegraphics[width=6cm]{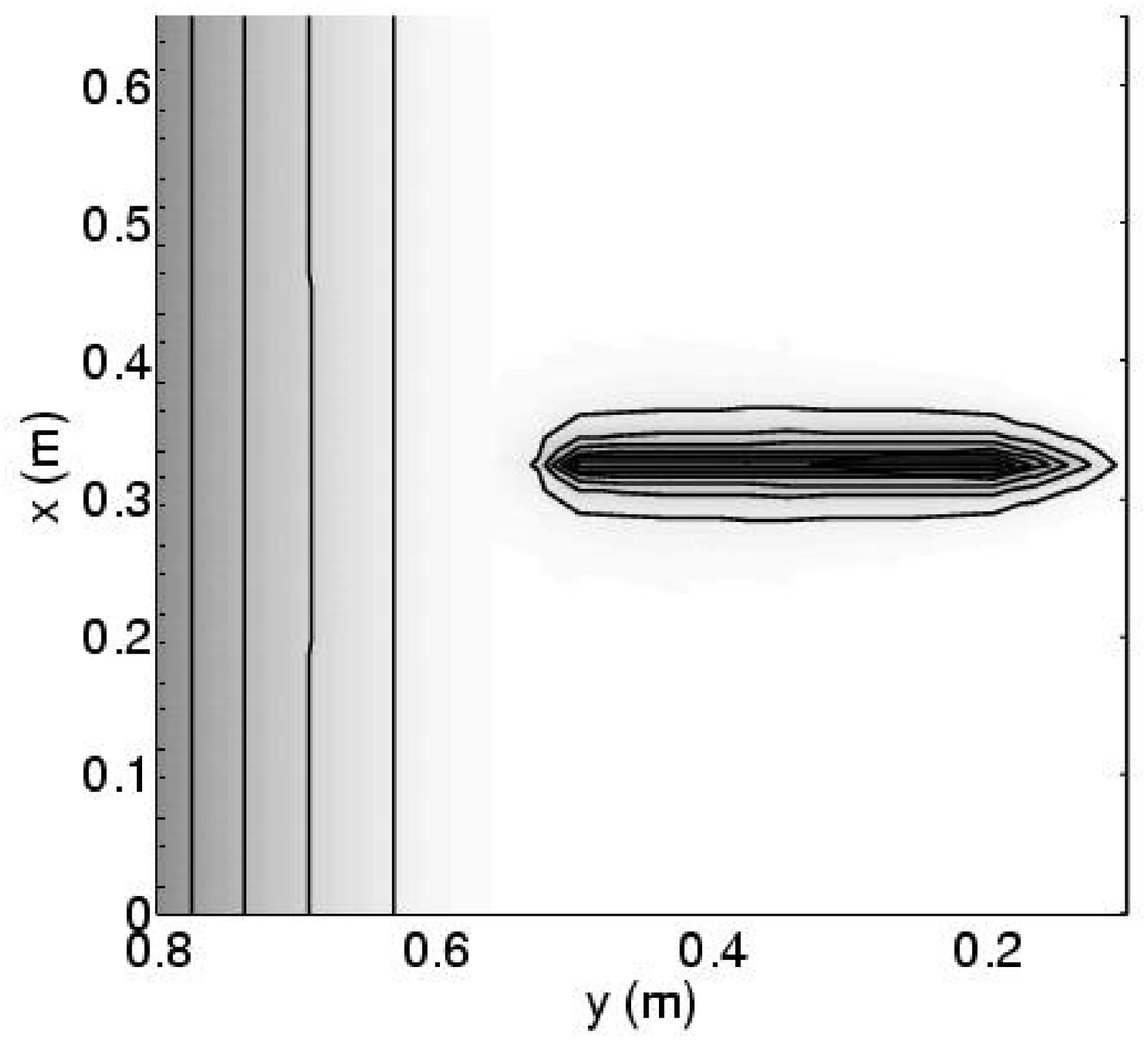}
\end{center}
\caption{The pressure distribution 1cm beneath the surface, on a plane
parallel to the slope.  Contour intervals are 10Pa.  A darker gray
tone indicates lower pressure.  a) The actual pressure distribution.
b) Pressure distribution when the lateral deformation of the water
table is neglected.  This case is unphysical, but illustrates that the
response of the water table is much shorter-reached.  In this
simulation the overall water table is first calculated without
channel.  Starting with this surface the channel is emplaced and the
pressure field calculated.  Only in (a) are resulting lateral
deformations of the water table taken into account.  The channel is
1cm deep (perpendicular to the $x-y$ plane), 1cm wide (parallel to the
$x$-axis), and 40cm long (parallel to the $y$-axis).
}\label{pressuredistr}
\end{figure}

Full numerical simulations bear out this effect in three dimensions.
Figure~\ref{pressuredistr}a shows the pressure in an inclined plane
1cm beneath the surface.  (Due to the atmospheric pressure boundary
condition, there are no pressure variations on the seepage surface
itself.)  The distance over which pressures decay corresponds to the
distance over which the velocity $v_x$ decays in
Figure~\ref{wentworth+birdview}b.  For the purpose of illustration,
Figure~\ref{pressuredistr}b shows the (unphysical) result if the lateral
deformation of the water table were neglected.  The lateral influence
of the channel is comparable to the channel width, while in
Figure~\ref{pressuredistr}a it is proportional to
the channel length.

\subsubsection{Coarsening}

The theoretical description is incomplete without quantitative
relations that describe the rate of erosion as a function of the
amount of water seeping out the surface and pouring down the channels.
Assuming that erosion is a monotonically increasing function of the
amount of water drawn into the channels, it is possible to predict the
dynamical behavior qualitatively.

Groundwater piracy should relate to channel spacing.  The
``wavelength'' $\lambda\approx c' \mathcal{W}$.  Here $c'$ is a
proportionality constant $c'\leq 1$, since piracy only gradually slows
the growth and does not instantly set the spacing.  According to
the previous discussion, $\mathcal{W}$ is proportional to $\ell$,
\begin{equation}
\mathcal{W}=c\mbox{ }\ell,
\label{mathcalW}
\end{equation}
where $c$ is a {\it geometric} constant, independent of permeability
or erosion threshold.  On the other hand, $c$ may depend on slope and
channel depth.
The resulting proposition is 
\begin{equation}
\lambda\approx c' c\hspace{0.2em}\ell.
\label{wavelength}
\end{equation}
Relation~(\ref{wavelength}) implies that seepage erosion has no
constant preferred wavelength, since the channel length is
time-dependent, $\ell=\ell(t)$.  This is consistent with the
observation of coarsening in the experiment.
Formula~(\ref{wavelength}) is not applicable to the early stage of
evolution, which is strongly influenced by the initial incision
processes.

Characteristic channel spacing can simply arise from the presence of a
dominant length scale in the longitudinal (flow) direction.  That is,
if the length of the channel is roughly predetermined, then the
drainage outlets would be spaced at a distance roughly proportional to
the length of the channels.  The Wentworth instability predicts that
channels keep on growing and competing, until the externally given
maximum length is reached.  In a natural setting the maximum length
may be determined by the length over which erosion occurs.

The scale-separation in the experiment may not be large enough to
distinguish whether spacing is proportional to channel width or
channel length.  However, in both theory and experiment competition
for ground water controls channel development.  The coarsening
mechanism is qualitatively the same.

In summary,
coarsening of the lateral inter-channel spacing is predicted by
theoretical considerations.  Channels compete for groundwater, which
slows and eventually stops the growth of some of them.  The
length-scale characterizing the distance between major channels
increases with time, until the channels have reached their maximum
length.  Eventually the lateral spacing should be only confined by the
externally constrained maximum length.

\section{Conclusions}

The lessons we have learned are about the role of runoff and seepage
during channel formation, the generic nature of rhythmic
channelization, and the importance of the water table's free surface.

We observe the formation of drainage outlet channels in permeable
ground, on the beach as well as in laboratory experiments.  In both
cases, initial channels are quickly incised by surface flow.  Even in
a very permeable medium, small amounts of surface runoff are more
efficient in starting channels than is the emerging groundwater.
Following the incision, groundwater from the vicinity seeps into the
channel and flows inside the channel, enlarging it, mainly lengthwise,
through erosion.

An initial lateral separation may be related to the initial sheet
flow.  Coarsening of the lateral inter-channel spacing is discernible
in the experiment.  For narrowly spaced channels, one channel often
slows and eventually stops its growth, presumably due to piracy of
groundwater by a nearby channel.  A second process taking place is the
belated formation of new channels far away from existing ones.
Rhythmic channel patterns arise as water seeps through a mono-disperse
permeable material and erodes material as it emerges and pours down
the channel.

These observations lead us to a theoretical description in terms of
flow of ground water in a homogeneous porous medium with a free water
table.  According to this theory, the response width of the water
table is much wider than the channel is wide or deep and approximately
proportional to the channel length.  This underscores Wentworth's
insight that this an efficient (and therefore the dominant)
instability mechanism, because it is prompt and far-reaching.
The groundwater flow pattern, for a given channel geometry,
is independent of material properties.

\vspace{3em}

We are indebted to many who have facilitated this research through
discussion, logistics, or help with field work, experiment, and
numerical simulation.  They include Oded Aharonson, Daniel Blair, Evan
Gould, Jack Holloway, Greg Huber, Andras and Monika Kalmar, Peter
Kelemen, Alexander Lobkovsky, Gary Parker, Taylor Perron, Antonello Provenzale, and
Kelin Whipple.  Peter Kelemen pointed us to the sites with regularly
spaced beach rills.  This work was supported by Division of Chemical
Sciences, Geosciences and Biosciences, Office of Basic Energy
Sciences, Office of Science, U.S. Department of Energy grants
DE-FG02-99ER15004 and DE-FG02-02ER15367.


\end{document}